\documentstyle[11pt]{article}

\begin{document}
\begin{flushright}
October 14, 1995
\end{flushright}

\vskip 10pt
\begin{center}
\Large\bf
Radio Science Investigation on \\
a Mercury Orbiter Mission
\vskip 26pt

\normalsize

J. D. Anderson$^1$, S. G. Turyshev$^1$, S. W. Asmar$^1$,  M. K. Bird$^2$,
A. S. Konopliv$^1$, \\
T. P. Krisher$^1$, E. L. Lau$^1$, G. Schubert$^3$ and W. L. Sjogren$^1$.

\vskip 8pt

$^1${\it Jet Propulsion Laboratory MS 301-230,
California Institute of Technology \\
 4800 Oak Grove Drive,
Pasadena, CA 91109-8099, USA.}
\vskip 8pt

$^2${\it Radioastronomisches Institut, Universit\"at Bonn\\Auf dem
H\"ugel 71, 53121 Bonn, Germany}
\vskip 8pt

$^3${\it Department of Earth and Space Sciences \\
Institute of Geophysics and Planetary Physics\\
University of California,  Los Angeles CA 90095-1567, USA}
\end{center}

\centerline{\bf Abstract}

We review  the results from {\it Mariner 10} regarding
Mercury's gravity field  and the results from radar ranging regarding
topography. We discuss the implications of improving these
results, including a determination of the polar component,
as well as the opportunity to perform  relativistic gravity tests with
a future {\it Mercury Orbiter}.
With a spacecraft placed in orbit with periherm
at 400 km altitude, apherm at 16,800 km, period 13.45 hr and latitude of
periherm at +30 deg, one can expect a significant improvement in our
knowledge of Mercury's gravity field  and geophysical properties.
The 2000 Plus mission that evolved during the European Space
Agency (ESA) {\it Mercury Orbiter} assessment
study (Hechler, 1994)  can   provide a global gravity
field complete through the 25th degree and order in spherical harmonics.
If after completion of the main mission, the periherm could be
lowered to 200 km altitude, the gravity field could be extended to
50th degree and order. We discuss the possibility that
a search for a Hermean ionosphere could be performed during the
mission phases featuring Earth occultations.

Because of its relatively large eccentricity and close proximity
to the Sun, Mercury's orbital motion provides one of the best
solar-system tests of general relativity.
Consequently, we emphasize the number of feasible
relativistic gravity tests that can be performed within
the context
of the parameterized post-Newtonian formalism - a useful
framework for testing  modern gravitational theories.
We pointed out that current results on relativistic precession
of Mercury's perihelion are uncertain by 0.5 \%, and we discuss
the expected improvement using {\it Mercury Orbiter}.
We discuss the importance of {\it Mercury Orbiter} for setting
limits on a possible time variation in the
gravitational constant $G$ as measured in atomic units.
Moreover, we mention that
by including a space-borne ultrastable crystal oscillator (USO)
or an atomic clock in the {\it Mercury Orbiter} payload, a new
test of the solar gravitational redshift would be possible
to an accuracy of one part in $10^4$ with a USO, and to an
accuracy of one part in $10^7$ with an atomic standard.  With an
atomic clock and additional hardware for a multi-link Doppler
system, including Doppler extraction on the spacecraft, the
effect of Mercury's gravity field on USO's frequency
could be measured with an accuracy of one part in $10^6$.
We discuss other relativistic  effects including
the geodetic precession of the orbiter's orbital plane
about Mercury, a planetary  test of the Equivalence Principle
(Nordtvedt effect), and  a
solar conjunction experiment to measure the relativistic time
delay (Shapiro effect).

\underline{Key Words:} Mercury, Mercury Orbiter, Radio Science,
Mercury gravity, Mercury ionosphere, Relativistic Gravitation
\vskip 1cm

\section{Introduction}

Mercury is the least explored of the terrestrial planets. Its
known global geophysical properties include its mass, radius, and
gravitational coefficients $J_2$ and $C_{22}$
(Anderson {\it et al.}, 1987). Recently, ground-based radar ranging
data and two range determinations from Mariner 10 have been used
to determine both the magnitude and direction of the equatorial
offset of the center of Mercury's figure from its center of mass,
and to determine the orientation and major and minor axes of the
ellipse which best fits Mercury's equatorial shape
(Anderson  {\it et al.}, 1995).
Displacements of the center of figure from the
center of mass are known for Earth (Balmino {\it et al.}, 1973), Moon
(Kaula  {\it et al.}, 1972; Bills and Ferrari, 1977; Smith {\it et al.},
1995), Venus (Bindschadler {\it et al.}, 1994), and Mars (Standish,
1973; Bills and Ferrari, 1978).  The global equatorial topography
of Mercury from ground-based radar ranging is available (Harmon
{\it et al.}, 1986; Harmon and Campbell, 1988; Pitjeva, 1993).  The
known shape and orientation of Mercury's equatorial figure and
the displacement of this figure from the planet's center of mass
place important constraints on the structure of Mercury's
interior.  The magnitude of the center of figure offset implies
an excess crustal thickness of 12~km or less, comparable to the
Moon's excess.  By comparing the equatorial ellipticity with the
Mariner 10 gravity coefficient C$_{22}$, and assuming Airy
isostatic compensation, Anderson  {\it et al.} (1995) conclude that
Mercury's crustal thickness is in the range 100 to 250~km.  A
similar calculation for the Moon, using $C_{22}=2.2\times
10^{-5}$ and $(a-b)/a=1.4\times 10^{-3}$, gives a mean crustal
thickness of about $\sim 72$ km (Schubert  {\it et al.}, 1995).
Because the radar ranging data are restricted between 12$^\circ$
and - 10$^\circ$ in latitude, {\it Mercury Orbiter} can provide
additional information on geophysical parameters aligned with the
polar axis.  It can also provide a global gravity field complete
through the 25th degree and order in spherical harmonics
(A polar orbit with periherm at   400 km altitude, apherm at
16,800 km, period = 13.45 hr, latitude of periherm at + 30 deg).
If after completion of the main mission, the periherm could be
lowered to 200 km altitude, the gravity field could be extended to
50th degree and order.

Radio propagation  measurements using spacecraft signals
have also provided scientific information on the atmospheres,
ionospheres, rings, tori, and surfaces of essentially every planetary
system encountered in the brief history of solar system
exploration (e.g. Tyler, 1987).
A search for a Hermean ionosphere could be performed with a
{\it Mercury Orbiter} during the
mission phases featuring Earth occultations.

Because of its relatively large eccentricity and close proximity
to the Sun, Mercury's orbital motion provides one of the best
solar-system tests of general relativity.  Its excess perihelion
precession, amounting to 42.98 arcseconds per century, is one of
the three classical tests of the theory.  Coherent Doppler and
ranging measurements using a {\it Mercury Orbiter} could provide
improved orbital information and hence, more sensitive tests.  A
number of predictions of alternative gravitational theories might
be detected as well, or failing a detection, tighter constraints
could be placed on permissible deviations from general
relativity.

By including a space-borne ultrastable crystal oscillator (USO)
or an atomic clock in the {\it Mercury Orbiter} payload, a new
test of the gravitational redshift would be possible.  The effect
of solar gravity on the oscillator's frequency could be measured
to an accuracy of one part in $10^4$ with a USO, and to an
accuracy of one part in $10^7$ with an atomic standard.  With an
atomic clock and additional hardware for a multi-link Doppler
system, including Doppler extraction on the spacecraft, the
effect of Mercury's gravity field could be measured with an
accuracy of one part in $10^6$.

This paper summarizes and discusses the different
radio science experiments proposed for
a future {\it Mercury Orbiter} mission.   The outline of the
paper is  as follows:  Section {\small II} discusses a present
 knowledge of Mercury's gravity
field and future improvements  offered by {\it Mercury Orbiter}.
In   section {\small III}, we  discuss  the possibility  of a
search for a Hermean ionosphere.
Mercury ranging data are reviewed in  section {\small IV}.
In section {\small V}, we discuss feasible
relativistic gravitational tests using the
satellite in orbit around Mercury.
In particular, we present the parameterized
post-Newtonian formalism, which has become a useful  theoretical
framework for analyzing the
gravitational experiments within the solar system.
This section contains a catalog of
relativistic gravitational experiments possible with a
{\it Mercury Orbiter}. We present  both the quantitative and
qualitative analyses of the measurable effects.
In  section {\small VI}, we  discuss briefly the
radio science instrumentation required for
the {\it Mercury Orbiter} mission under consideration by ESA.

\section{Mercury Gravity Field}

The determination of Mercury's gravity field could be obtained by
the  reduction of Doppler radio tracking data.  This is an old
and proven technique used for the moon, Mars and Venus (Konopliv and
Sjogren, 1994).  The field would be modeled with spherical harmonic
coefficients does providing a smooth global model.  Given
the power of modern computers, it is feasible to estimate
literally thousands of harmonic coefficients.  The resulting
coefficients would be interpreted in terms of a gravity surface
function, which would be plotted as a surface gravity map.  For
other planets, in particular Venus, Earth, the moon and Mars, the
correlation of gravity maps with surface topography and imaging
has revealed the  isostatic state of surface features.  For
example, similar features at different locations on a planet can
have significantly different internal density variations or
structure.  Geophysical interpretation can yield information on
the distribution of crustal  thickness and dynamic hot spots.
The degree of isostatic compensation may indicate an old crust,
nearly completely compensated, or a crust with sufficient
rigidity to imply a cold interior.

For Mercury, the second order gravity
coefficients $C_{20}, C_{21}, S_{21}, C_{22}$ and $S_{22}$ would provide the
orientation  of the axes of the moments of inertia, which would
be compared with the spin axis orientation.  A more accurate $C_{20}$
coefficient would provide a better-determined gravitational oblateness,
and  $C_{22}$  coefficient would reveal  a better-determined  gravitational
ellipticity of the equator.  At a more local level, all the other
coefficients, to as high a degree and order as possible, would
provide a gravity map for comparison with surface topography.
The most obvious and interesting correlations would be negative
(i.e., gravity highs in topographic lows or gravity lows in
topographic highs).   On the moon and Mars there are negative
correlations for some large basins  (but not all).  This has
placed strong constraints on interior thermal models.   We
anticipate that a gravity map of Mercury's Coloris Basin would
place significant constraints on its underlying internal
structure.

As a rule of thumb, the spatial resolution of gravity features
derived from  coherent X-band  Doppler data would be less than
the spacecraft altitude.  For a 400 km  altitude periherm, the
best resolution would be approximately  300 km or $7.1^\circ$ on
the surface of Mercury and would correspond to  a 25th degree and
order spherical harmonic model.  At a lower 200 km altitude a
50th degree and order model could be obtained for a total of 2600
harmonic coefficients. The ideal data acquisition period would be
59  days of Doppler data covering at least $\pm1$  hour of
periapsis. Data within $5^\circ$ to $10^\circ$
of solar conjunction would be too noisy for gravity field
determination. During gravity passes other
non-gravitational  forces would be minimized.  These data
would provide complete global  coverage.  Data over 59 days at a
different viewing geometry would  significantly strengthen the
solution.

In addition to the gravity coefficients,  other parameters would
be estimated simultaneously from the coherent Doppler data.
These include the Love number $k_2$, which would provide a
measure of whether there is a solid or liquid core,
translational forces from solar radiation
impinging on spacecraft parts, Mercury's albedo,  corrections
to the Mercury ephemeris, and location of the spin pole
and rotation rate.

The final data products that would be produced for geophysical
interpretation are:

1.  The spherical harmonic coefficients.

2.   A full covariance matrix on all the coefficients (i.e.,
$1-\sigma$ uncertainties and their

\hskip 0.5cm correlations).

3.  A $1^\circ \times 1^\circ$ digitized vertical surface
gravity map.

4. A $1^\circ \times 1^\circ$ digitized geoid map.

5. $1-\sigma$ uncertainty maps on gravity and geoid.

6. If topography data, $1^\circ \times 1^\circ$  Bouguer
gravity maps.

7. $1^\circ \times 1^\circ$ Isostatic Anomaly map.

\section{Mercury Ionosphere}
Mercury  was encountered three times by the spacecraft {\it
Mariner 10}, but was occulted by the planet's disk only at the
first flyby on 29 March 1974. The S/X-band dual-frequency radio
links from spacecraft to Earth were monitored during the flyby
(Howard  {\it et al.}, 1974). The attempt to detect the ionosphere
from the dispersive phase shift between the two downlinks,
however, yielded only an upper limit of 1000 cm$^{-3}$ for the
electron density in the altitude range from the surface up to
1000 km (Fjeldbo {\it et al.}, 1976). These observations were
used to infer an upper limit to the surface density of the
Hermean atmosphere of 10$^{6}$ molecules cm$^{-3}$. Taking a
surface temperature of 500 K, this corresponds to a pressure of
ca.\ 10$^{-13}$ bar (a very decent vacuum!).  The mission profile
that evolved during the ESA {\it Mercury Orbiter} assessment
study (Hechler, 1994) offers definite prospects for radiometric
investigations. The geometry has many parallels to that designed
for {\it Mars Orbiter}, i.e. regularly recurring occultations at
polar latitudes during certain phases of the mission (Tyler {\it
et al.}, 1992). Whereas the atmosphere on Mars is dense enough to
be synoptically measured on a global basis, Mercury's high
temperature and low surface gravity combine to produce only a
residual ``exosphere'' with the above mentioned diminutive
surface densities.  Detecting an ionosphere on Mercury using
radio occultation techniques is a difficult, but not hopeless,
task. Generally speaking, the ionosphere should be easier to
detect than the neutral atmosphere. One classical example of an
ionospheric detection was the {\it Pioneer 10} measurement during
occultation of the Jovian moon Io (Kliore {\it et al.}, 1974),
for which peak densities of 6 $\times$ 10$^{4}$ electrons
cm$^{-3}$ were derived. The geometry of this particular
occultation enabled the detection to be made with only an S-band
downlink. The ionospheric effect on the S-band downlink during
the {\it Voyager~2} occultation of Neptune's moon Triton was
about five times greater than that of the neutral atmosphere
(Tyler {\it et al.}, 1989). There are, of course, counterexamples
such as Titan (Lindal {\it et al.}, 1983), which produced
dramatic refraction effects from its dense, extended atmosphere,
but no discernable ionospheric signature.  Radio propagation
experiments on previous ESA interplanetary missions (e.g.\  {\it
Giotto}, {\it Ulysses}) have utilized the extensive facilities of
the NASA Deep Space Network. It is expected that similar ground
equipment will be available for the {\it Mercury Orbiter}. The
assessment study for the {\it Mercury Orbiter} recommended an
S/X-Band radio communications system for both up- and downlinks
(ESA Study Team, 1994). For the dual-frequency (S/X-band)
downlink configuration on {\it Ulysses}, the smallest detectable
change in differential phase (i.e.\ S-band minus 3/11 X-band) was
estimated to be ca.\ 5$^{\circ}$ at S-band, corresponding to an
electron content of about 10$^{14}$ electrons m$^{-2}$ (Bird {\it
et al.}, 1992). The occultation geometry afforded by an orbiting
spacecraft, however, for which the lateral motion of the signal
ray path through the ionized medium is quite rapid, may still
enable detection of intervening plasma without the dual-frequency
capability.  For a geometry of this type, the smallest phase
change measureable is estimated as $\sim$0.5$^{\circ}$. This is
also the level of detectability attributed to with the {\it
uplink} radio occultation instrumentation presently being
developed for the {\it Pluto Express} Mission [G.L. Tyler, 1995,
private communication]. This tiny change in phase, which provides
a fair estimate of the present-day detection threshold,
corresponds to a neutral column density of 5 $\times$ 10$^{24}$
molecules m$^{-2}$ or an electron columnar content of 10$^{14}$
electrons m$^{-2}$ at X-band. For a spherically symmetric Hermean
atmosphere (ionosphere), this can be shown to translate to peak
number densities along the ray path at the planet's limb of
5~$\times$~10$^{12}$ molecules cm$^{-3}$ (100 electrons
cm$^{-3}$). Using the upper limits quoted by Fjeldbo
{\it et al.} (1976), column densities for the ray path tangent to the limb
may be estimated for Mercury. The neutral column density is found
to be 10$^{18}$ molecules m$^{-2}$, which is, unfortunately,
about six orders of magnitude below the detection threshold.
Phase changes induced by the neutral gas are thus probably
negligible.  The ionized component, however, if we assume the
peak reaches 10$^3$ electrons cm$^{-3}$, can produce a content of
10$^{15}$ electrons m$^{-2}$, which would place it above the
lower limits of detectability. This estimate, it should be
emphasized, is only valid for a global-scale ionosphere. However,
as pointed out during the {\it Mercury Orbiter} assessment study
(ESA Study Team, 1994), atomic sodium, which emanates from
various hot spots on the planet's surface, may well be the most
prevalent component of the Hermean exosphere. Mean surface
densities of up to 5~$\times$~10$^{4}$~cm$^{-3}$ are inferred
(Cheng {\it et al.}, 1987). The Na atom is photoionised on time
scales of a few hours, so this is presumed to be the most
important source of ionospheric plasma. The proximity of the
magnetopause and the tailward-streaming solar wind, however, may
cause many ions to be swept away at rates even quicker than those
of the sources. If the electron density is concentrated in the
atmosphere above a local source, densities considerably higher
than 1000~cm$^{-3}$ would be necessary to produce a measureable
effect.  Regularly recurring Earth occultations, during which
radio tracking at ground-based stations is interrupted, are
included in virtually every {\it Mercury Orbiter} mission
scenario because they are basically ``unavoidable''. Two such
occultation intervals, one near inferior conjunction and one near
superior conjunction, were found to occur for the prototype {\it
Mercury Orbiter} mission (Hechler, 1994). The total number of
such occultations, each with ingress and egress opportunities, is
well over 100.

Another feature, as mentioned above, is the rapid swath taken by
the radio ray path through the ionosphere. This magnifies the
plasma Doppler shift on the occulted radio signal, thereby
increasing the detection probability.  Synoptic Doppler
observations of ionospheric electron densities on a {\it Mercury
Orbiter} would thus appear to be very difficult, but still worth
a good look. Even in the absence of a permanent global ionosphere
it is still possible, however, that a systematic examination of
the radio occultation data would result in the serendipitous
detection of near-surface plasma clouds.

\section{Mercury Ranging Data}
 A summary and
discussion of data used in the fundamental ephemerides
DE200/LE200 is available (Standish, 1990; Standish  {\it et al.}, 1992).
Those data include Mercury radar ranging data spanning the years
1966-1974.  Analysis of the older data plus additional data
spanning the years 1974 to 1990 is underway (Anderson et al.,
1995).  Included are two range fixes from the {\it Mariner 10} Mercury
flybys on March 29, 1974 and March 16, 1975 (Anderson  {\it et al.},
1987).  The current JPL set of reduced Mercury radar ranging data
is summarized in Table 1.  The Goldstone data are from 34 m and
70 m stations located in California's Mojave desert, the Arecibo
data are from Puerto Rico, while the Haystack data are from
Tyngsboro, Massachusetts.

The data used in current relativity tests are reduced ranging
measurements to Mercury's surface.  Both radar time delay and
Doppler data have been used in the reduction (for an explanation
of the Doppler-delay technique see e.g. Ingalls and Rainville
1972; Shapiro  {\it et al.}, 1972; Harmon  {\it et al.}, 1986).
The 1978-1982
reduced data are from archives at the Harvard Smithsonian Center
for Astrophysics (J. Chandler, private communication). Systematic
errors in radar ranging data limit their effectiveness for tests
of general relativity, most notably the excess precession of
Mercury's perihelion, and topography models have been introduced
for purposes of minimizing the error (Anderson {\it et al.}, 1991;
Pitjeva, 1993; Anderson {\it et al.}, 1995).  New data, much less
affected by systematic error, will become available with {\it Mercury
Orbiter}.  Specifically, transponded ranging, similar to that from
the {\it Mariner 9 Mars Orbiter}, will be available. Eventually, with
the introduction of a {\it Mercury Lander}, even better surface
transponded ranging will become available, similar to ranging
data from the two {\it Viking Landers}.  Unfortunately, no {\it Venus
Orbiter} has carried a ranging transponder to date. However, we
have removed Venus topography from Venus radar ranging data by
using the {\it Pioneer 12} radar altimetry measurements, and one range
fix is available from the {\it Galileo} spacecraft flyby in 1990
(Anderson  {\it et al.}, 1991).   In addition, the MGN X-band Doppler
data provide a measure of the  line-of-sight velocity to an
accuracy of 0.02 mm/s  (or $\approx$ 10m range accuracy). Of all
the inner planets, only Mercury currently requires a
parameterized topography model for ranging data analysis.  The
{\it Mercury Orbiter} will remove this restriction, and the orbital
motion of Mercury will be placed on an equal footing with Venus
and Mars.

\section{Relativistic Gravitational Tests}

Mercury is the closest to the Sun of all the planets of the
terrestrial group  and,
due to its unique location and orbital parameters, it is
especially suited to relativistic
gravitational experiments. The short period of its solar orbit
will allow experiments over several orbital revolutions.
 In this section  we will   define the list of gravitational
expermients  for the {\it Mercury Orbiter} mission.
Depending on the nature of the measureable effect, there are two main
groups of possible  experiments, namely:
(i) the celestial mechanical studies of the
orbital motion  and rotation of
Mercury in the gravitational field of the Sun,
as well as  observations of the
motion of the {\it Mercury Orbiter} in the solar
gravitational field at the vicinity of the
planet Mercury, and (ii) the  studies of
the gravitational influence on the
propagation of the radio waves  along the  Earth-Mercury path.
We will discuss separately   the possible
experiments from both these
groups and will present  the estimates of the magnitude and  physical
character of the measured effects.
Analysis, performed in this section, is directed
towards the future mission, so we
will show which relativistic effects can be measured, and how
accurately  these measurements can be done.

\subsection{Parameterized Post-Newtonian Gravity.}

Metric theories of gravity have a special position
among all the other possible theoretical models.
The reason  is that, independently of the many different
principles at their foundations, the gravitational field
in these theories affects the matter directly through the metric
tensor of the Riemannian space-time   $g_{mn}$, which is
determined from the field equations of a particular theory of gravity.
As a result, in contrast to Newtonian gravity, this tensor
contains the properties of a particular gravitational theory as well as
carries the information about the gravitational field of the bodies.
This property of the metric tensor enables one to analyze
the motion of matter in one or another metric theory of
gravity  based only on the basic principles
of modern theoretical physics.

In order to accumulate the features of many modern metric theories of
gravity in one  theoretical scheme as well as
to create the `versalite instrument' to plan  the gravitational
experiments and to analyze the
data obtained,  Nordtvedt and Will have proposed
the parameterized post-Newtonian
({\small PPN}) formalism (Nordtvedt 1968a;
Will  1971; Will \& Nordtvedt 1972).
This formalism allows one to describe within the
common framework the motion of
the celestial bodies  for a wide class of metric
theories of gravity.
Within the accuracy of modern experimental
techniques, the {\small PPN} formalism
provides a useful foundation for testing the
 predictions of different metric
theories of gravity within the weak gravitational field and
slow motion approximation appropriate for the solar system.

 The metric tensor of the general Riemannian
space-time in the {\small PPN}
formalism  is generated by some given distribution of matter in the
form of an ideal fluid. It is represented by a sum of
gravitational potentials with   arbitrary coefficients,  the
 {\small PPN} parameters. The most general form of this metric tensor
in four dimensions   may be written
as follows\footnote{The geometrical units $\hbar=c=G=1$ are used
throughout  as is the metric convention $(+---)$.}:
{}
$$g_{00}= 1-2U+2(\beta+\tau) U^2-(2\gamma+2+\alpha_3+\zeta_1)\Phi_1+
\zeta_1 A+2(\zeta+\tau)\Phi_w-$$
{}
$$-2\Big[(3\gamma+1-2\beta+\tau+\zeta_2)\Phi_2+
(1+\zeta_3)\Phi_3+3(\gamma+\zeta_4)\Phi_4+
\nu  \chi,_{00} \Big]-$$
{}
$$-(\alpha_1-\alpha_2-\alpha_3) w_\mu w^\mu U+
\alpha_2 w^\mu w^\nu U_{\mu\nu}+(\alpha_1-2\alpha_3)w^\mu V_\mu+
{\cal O}(c^{-6}), \eqno(1a)$$
{}
$$g_{0\alpha}={1\over 2}(4\gamma+3+2\nu+2\tau+
\alpha_1-\alpha_2+\zeta_1)V_\alpha+$$
{}
$$+ {1\over2}(1-2\nu-2\tau+\alpha_2-\zeta_1)W_\alpha-
{1\over2}(\alpha_1-2\alpha_2)w_\alpha U+\alpha_2 w^\mu U_{\alpha\mu}
+  {\cal O}(c^{-5}),\eqno(1b)$$
{}
$$g_{\alpha\beta}=\gamma_{\alpha\beta}(1+2(\gamma+\tau) U)  +
2\tau U_{\alpha\beta}+  {\cal O}(c^{-4}).\eqno(1c)$$

\noindent with the generalized gravitational potentials given
as follows:
{}
$$U(z^p)  = \int {\rho_0 (z'^p)\over |z^\nu - z'^\nu|} d^3z'^\nu,
\qquad V^\alpha(z^p)  = - \int{\rho_0(z'^p)v^\alpha (z'^p)\over
|z^\nu-z'^\nu|}d^3z'^\nu,$$
{}
$$W^\alpha(z^p)  = \int\rho_0(z'^p)v_\mu(z'^p)
{(z^\alpha-z'^\alpha)(z^\mu-z'^\mu)\over|z^\nu-z'^\nu|}d^3z'^\nu,$$
{}
$$A(z^p)  = \int\rho_0(z'^p){ [v^\mu(z'^p)(z^\mu-z'^\mu)]^2
\over|z^\nu-z'^\nu|^3}d^3z'^\nu,
\qquad \chi(z^p)  = \int  \rho_0 (z'^p)
|z^\nu - z'^\nu|  d^3z'^\nu, $$
{}
$$U^{\alpha\beta}(z^p)  = \int\rho_0(z'^p){(z^\alpha-z'^\alpha)
(z^\beta-z'^\beta)\over|z^\nu-z'^\nu|^3}d^3z'^\nu,$$
{}
$$\Phi_1(z^p)  = - \int {\rho_0(z'^p) v_\lambda
(z'^p)v^\lambda(z'^p)\over
|z^\nu  - z'^\nu|}d^3z'^\nu,\qquad \Phi_{2} (z^p)
= \int{\rho_0(z'^p)U(z'^p)\over
|z^\nu - z'^\nu |}d^3z'^\nu, $$
{}
$$\Phi_{3} (z^p) =  \int { \rho_0(z'^p)\Pi(z'^p)\over
|z^\nu  -z'^\nu|}d^3z'^\nu, \qquad \Phi_{4} (z^p) =
\int{\rho_0(z'^p)  p(\rho(z'^p))\over |z^\nu- z'^\nu|}
d^3z'^\nu,$$
{}
$$\Phi_w(z^p)  = \int\int\rho_0(z'^p)\rho_0(z''^p)
{(z^\beta-z'^\beta)\over|z^\nu-z'^\nu|^3}
\Big[{(z_\beta-z''^\beta)\over|z'^\nu-z''^\nu|}
-{(z'^\beta-z''^\beta)\over|z^\nu-z''^\nu|}
\Big]d^3z'^\nu d^3z''^\nu. \eqno(2)$$

Besides the   {\small PPN} parameters
$(\gamma,\beta,\zeta,\alpha_1-\alpha_3;\zeta_1-\zeta_4)$,
the expression
(1) contains  two other parameters  $\nu$ and $\tau$
(Denisov and Turyshev, 1990; Turyshev, 1990).
The  parameter $\nu$ reflects the specific
choice of the gauge conditions and for the standard
{\small PPN} gauge it is  given as  $\nu=0$, but for harmonic gauge
conditions one should choose $\nu={1\over2}$. The parameter $\tau$
describes  the possible anisotropy of the space-time and it
corresponds to a different  spatial coordinates,
 which may be  chosen for
modelling the experimental situation.
For example, the case $\tau=0$ corresponds to
harmonic coordinates, while $\tau=-1$
corresponds to the standard (Schwarzschild)
coordinates.
Then, a particular metric theory of gravity in the {\small PPN}
formalism with a  specific coordinate
gauge $(\nu,\tau)$ might be fully characterized by the means of
the ten {\small PPN} parameters.
This formalism uniquely prescribes the values
of these parameters for each particular theory under study.
In the  standard {\small PPN} gauge
(i.e. in the case when $\nu=\tau=0$) these  parameters have clear
physical meaning.
The parameter $\gamma$ represents the
measure of the curvature of the space-time
created by the unit rest mass; the parameter $\beta$ is the measure of the
non-linearity of the law of superposition of
 the gravitational fields in a
theory of gravity (or the measure of the metricity);
the parameter $\zeta$ quantifies the
violation of the local position
invariancy; the group of parameters $\alpha_1, \alpha_2, \alpha_3$
specify the violation of  Lorentz invariancy  (or the presence of the
privileged reference frame), and, finally, the parameters
$\zeta_1,\zeta_2,\zeta_3,\zeta_4$ reflect  the violation of the
law of total momentum conservation for a closed gravitating system.
Note that general relativity, when analyzed in  standard
{\small PPN} gauge, gives: $\gamma=\beta=1$ and all the other
eight parameters vanish.
Whereas for  the Brans-Dicke theory,
$\beta=1, \gamma= {1+\omega\over 2+\omega}$, where $\omega$
is an unknown dimensionless parameter of this theory.

 The main properties  of the solution (1)-(2) are well
established and widely in use in modern
astronomical practice (Moyer, 1971; Moyer, 1981; Brumberg, 1991;
Standish {\it et al.}, 1992;  Will, 1993).
For  practical purposes one   chooses the inertial
reference frame located in the center of mass of an
isolated distribution of matter,
 then, by performing a power expansion of the potentials in terms of
spherical harmonics, one obtains the
post-Newtonian  definitions of the mass of  the
body  $m_a$, it's center of mass ${\cal Z}^{\alpha}_a$,
momentum ${\cal P}^{\alpha}_a$ and angular momentum
$S^{\alpha\beta}_a$. Thus the definitions for the mass
$m_a$ and coordinates of the center of mass of the
 body ${\cal Z}^{\alpha}_a$  in any inertial
reference frame are given by the formulae (for more
detailed analysis see (Damour, 1983) and (Will,  1993)
 and references therein):
{}
$$m_a = \int_a d^3 z'^{\epsilon }_a  \hskip 1mm
 \rho^*_{a0}  \Big( 1 +\Pi  + {1\over2}  U -
 {1\over2}v_\lambda v^\lambda
+{\cal O}(c^{-4})  \Big),\eqno(3a)$$
{}
$${\cal Z}^{\alpha}_a(t) =  {1\over m_a }  \int_a d^3 z'^{\epsilon }_a
\hskip 1mm    \rho^*_{a0}   z'^{\alpha}_a
\Big( 1 +\Pi  + {1\over2}  U -   {1\over2}v_\lambda v^\lambda
+{\cal O}(c^{-4}) \Big). \eqno(3b)$$

\noindent where the conserved mass density is defined by
$\rho^*_{a0}={\rho_a}_0 \sqrt{-g} u^0$.

It has been shown (Fock, 1959; Ni \& Zimmerman, 1978)
 that for an  isolated distribution of matter in this approximation
 there exist a set of inertial reference frames  and ten integrals
of motion  corresponding to ten conservation laws.
Moreover, Ni and  Zimmerman have shown (1978) that in order for a metric
theory to have the total energy-momentum conservation laws
(so-called fully conservative theories),
 only three {\small PPN} parameters  may have non-zero means, namely
$\gamma, \beta$ and $\zeta$. For the future analysis of
gravitational experiments on
{\it Mercury Orbiter},  we will limit ourselves to this class of  metric
theories of gravity.

 In the post-Newtonian approximation
the mass $m_a$ (3a) is conserved and the centre of  mass
${\cal Z}^{\alpha}_a$  moves in
space with a constant velocity along a straight line.
Thus, ${\cal Z}^\alpha_a(t) =
{\cal P}^\alpha \cdot t + {\cal K}^\alpha$,
where the constants $ {\cal P}^\alpha =
 d {\cal Z}^{\alpha}_B /d t$
and ${\cal K}^\alpha$ are the body's momentum and center of
inertia, respectively. One can choose from the set of
inertial reference frame the barycentric inertial
 one. In this frame the functions
${\cal Z}^\alpha$ must equal zero for any moment of time.
This condition can be satisfied by applying to the metric
eqs.(1) the post-Galilean transformations
(Chandrasekhar \& Contopulos, 1967)
 where  the constant velocity $u^\alpha$
 and displacement of origin $b^\alpha $
should be selected in a such a way that
${\cal P}^\alpha$ and ${\cal K}^\alpha$
equal zero (for details see
 Kopejkin, 1988; Will, 1993). The resulting
barycentric inertial reference frame  has been
adopted for the fundamental
planetary and lunar ephemerides (Newhall {\it et al}., 1983;
Standish {\it et al.}, 1992). Moreover, the
coordinate time of
the solar barycentric (harmonic) reference frame
is the {\small TDB} time scale,
which has been adopted in modern astronomical
practice (Fukushima, 1995).

In order to analyze the  motion of bodies  in the solar system
barycentric reference frame  one may obtain the Lagrangian function $L_N$
describing the motion of {\small N} extended self-graviting
bodied (Turyshev, 1990; Will, 1993).
Within the accuracy necessary for our future discussion of the
gravitational experiments on
the {\it Mercury Orbiter} mission, this function  can be presented
in the following form:
{}
$$L_N=\sum_a^N{m_a\over 2}{v_a}_\mu{v_a}^\mu
\Big(1-{1\over 4}{v_a}_\mu{v_a}^\mu\Big)-$$
{}
$$- \sum_a^N\sum_{b\not=a}^N{m_a m_b\over r_{ab}}\Bigg({1\over2}
+(3+\gamma-4\beta +3\zeta)\Omega_a-
\zeta {n_{ab}}_\lambda {n_{ab}}_\mu\Omega_a^{\lambda\mu}-$$
{}
$$-(\gamma+\tau+{1\over 2}){v_a}_\mu {v_a}^\mu +
(\gamma+\tau+{3\over 4}){v_a}_\mu {v_b}^\mu- $$
{}
$$ -({1\over 4}-\tau) {n_{ab}}_\lambda {n_{ab}}_\mu
{v_a}^\lambda {v_b}^\mu -\tau ({n_{ab}}_\mu  {v_a}^\mu)^2+$$
{}
$$+{{n_{ab}}_\lambda\over r_{ab}} \Big[(\gamma+{1\over 2}){v_a}_\mu-
 (\gamma+1){v_b}_\mu\Big] S_a^{\mu\lambda} + {n_{ab}}_\lambda
{n_{ab}}_\mu {I_a^{\lambda\mu}\over r_{ab}^2}\Bigg)+$$
{}
$$+(\beta+\tau-{1\over 2}) \sum_a^N m_a\Big(
\sum_{b\not=a}^N {m_b\over r_{ab}}\Big)^2+ $$
{}
$$+ (\zeta+\tau)\sum_a^N\sum_{b\not=a}^N \sum_{c\not=a,b}^N
m_a m_bm_c\Big[ {{n_{ab}}_\lambda\over 2r_{ab}^2}
(n_{bc}^\lambda+n_{ca}^\lambda)-{1\over r_{ab}r_{ac}}\Big]+
\sum_a^N m_a {\cal O}(c^{-6}), \eqno(4) $$

\noindent  where $m_a$ is the isolated rest mass of a body $a$,
the vector $r_a^\alpha$ is the barycentric radius-vector of this
body, the vector $r_{ab}^\alpha=r_b^\alpha -r_a^\alpha$ is the
vector directed from body $a$ to body  $b$, and
the vector $n^\alpha_{ab}=r^\alpha_{ab}/r_{ab}$ is the usual notation
for the unit vector along
this direction. It should be noted that the expression (4)
does not depend on the parameter $\nu$, which confirms that
this parameter is the gauge parameter only.
The tensor $I_a^{\mu\nu}$ is the spatial
trace-free ({\small STF}) (Thorne, 1980)
tensor of the quadrupole moment of
body $a$  defined by
{}
$$I_a^{\mu\nu}={1\over 2m_a}\int_a d^3z'^\nu_a\rho^*_{a0}(z'^p_a)
\Big(3z'^\mu_a z'^\nu_a-\gamma^{\mu\nu}z'_{a\beta} z'^\beta_a \Big)=
3{J_a}^{\mu\nu}_2-\gamma^{\mu\nu}{J_a}_2, \eqno(5)$$

\noindent where the quantity ${J_a}_2$ is
the  quadrupole coefficient. The tensor $S_a^{\mu\nu}$
is the body's intrinsic  {\small STF} spin moment  which is given as:
{}
$$S_a^{\mu\nu}={1\over m_a}\int_a  d^3z'^\nu_a \rho^*_{a0} (z'^p_a)
\big[u_a^\mu z'^\nu_a-u_a^\nu z'^\mu_a\big], \eqno(6)$$

\noindent where $u_a^\mu$ is the rotational
velocity of body $a$. Finally, the tensor $\Omega_a^{\mu\nu}$
is the body's gravitational binding energy tensor:
{}
$$\Omega_a^{\alpha\beta} = -{1\over 2m_a}{\int\int}_a
d^3z'^\nu_a d^3z''^\nu_a \rho_0(z'^p_a)\rho_0(z''^p_a)
{(z'^\alpha_a-z''^\alpha_a)(z'^\beta_a-z''^\beta_a)
\over|z'^\nu_a-z''^\nu_a|^3}. \eqno(7)$$

\noindent Furthermore, with the help of the Lagrangian  (4),
one can obtain the corresponding equations of motion of an
arbitrary extended body $a$ as follows:
{}
$$\ddot{r}^\alpha_a =\sum_{b\not=a}
{{{\cal M}_b}^\alpha_\beta \over r_{ab}^2}{\widehat{n}_{ab}}^\beta +
\sum_{b\not=a} {m_b\over r_{ab}^2}\Big[{\cal A}^\alpha_{ab}+
{{{\cal B}^\alpha_{ab}}\over r_{ab}}+
{{\cal C}^\alpha_{ab} \over r_{ab}^2}-$$
{}
$$-{n_{ab}^\alpha\over r_{ab}}\Big((2\beta+2\gamma+
2\tau+1)m_a+(2\beta+2\gamma+2\tau)m_b\Big)\Big]+$$
{}
$$+ \sum_{b\not=a}\sum_{c\not=a,b}m_bm_c
{\cal D}^\alpha_{abc}+{\cal O}(c^{-6}), \eqno(8)$$

\noindent where, in order to account for the influence of the
gravitational binding energy $\Omega^{\alpha\beta}_b$,
we have introduced in  the equation (8)
the tensor of passive gravitational rest mass $ {\cal M}^{\alpha\beta}_b$
  as follows
{}
$${\cal M}^{\alpha\beta}_b = m_b\Big[\delta^{\alpha\beta}
\Big(1+(3+\gamma-4\beta+3\zeta)\Omega_b-
3\zeta {n_{ab}}_\lambda {n_{ab}}_\mu\Omega^{\lambda\mu}_b
\Big)+2\zeta \Omega^{\alpha\beta}_b +{\cal O}(c^{-4})\Big], \eqno(9)$$

\noindent and the unit vector  $n_{ab}$ has also been
corrected by the gravitational binding energy
and the  tensor of the quadrupole moment
$I^{\alpha\beta}_a$ of the body $a$
under question:
{}
$$\widehat{n}^{\alpha}_{ab} = n^{\alpha}_{ab}\Big(1+(3+\gamma-4\beta+
3\zeta)\Omega_a-
3\zeta {n_{ab}}_\lambda {n_{ab}}_\mu\Omega^{\lambda\mu}_a+
5{n_{ab}}_\lambda {n_{ab}}_\mu
{I^{\lambda\mu}_a\over r_{ab}^2}\Big)+$$
{}
$$+ 2\zeta {n_{ab}}_\beta\Omega^{\alpha\beta}_a +
2{n_{ab}}_\beta {I^{\alpha\beta}_a\over r_{ab}^2}+{\cal O}(c^{-4}).
\eqno(10)$$

\noindent The term ${\cal A}^\alpha_{ab}$ in the expression (8)
is the orbital term  which is given as follows:
{}
$${\cal A}^\alpha_{ab}= v^\alpha_{ab} {n_{ab}}_\lambda\Big(v^\lambda_a-
(2\gamma+2\tau+1)  v^\lambda_{ab}\Big)+$$
{}
$$+ n^\alpha_{ab}\Big({v_a}_\lambda v^\lambda_a-
(\gamma+1-\tau) {v_{ab}}_\lambda v^\lambda_{ab}+
3\tau ({n_{ab}}_\lambda {v_{ab}}^\lambda)^2-
{3\over 2}({n_{ab}}_\lambda v^\beta_{b})^2\Big), \eqno(11)$$

\noindent The spin-orbital term $ {\cal B}^\alpha_{ab} $ has the form:
{}
$$ {\cal B}^\alpha_{ab} =({3\over 2}+2\gamma){v_{ab}}_\lambda
(S^{\alpha\lambda}_a+S^{\alpha\lambda}_b)+{1\over2}{v_a}_\lambda
(S^{\alpha\lambda}_a-S^{\alpha\lambda}_b)+$$
{}
$$ +{3\over 2}(1+2\gamma){n_{ab}}_\lambda{v_{ab}}_\beta\Big[n_{ab}^\beta
(S^{\alpha\lambda}_a+S^{\alpha\lambda}_b)-{n_{ab}}^\alpha
(S^{\beta\lambda}_a+S^{\beta\lambda}_b)\Big]+$$
{}
$$+{3\over 2}{n_{ab}}_\lambda \Big[n_{ab}^\alpha
({v_{a}}_\beta S^{\beta\lambda}_b-{v_{b}}_\beta S^{\beta\lambda}_a) +
{n_{ab}}_\beta{v_{ab}}^\beta S^{\alpha\lambda}_b\Big]. \eqno(12)$$

\noindent  The term ${\cal C}^\alpha_{ab}$ is
caused by the oblateness of the bodies in the system:
{}
$$ {\cal C}^\alpha_{ab} =2{n_{ab}}_\beta I^{\alpha\beta}_b +
5n_{ab}^\alpha{n_{ab}}_\lambda {n_{ab}}_\mu I^{\lambda\mu}_b.
 \eqno(13) $$

\noindent And, finally, the contribution ${\cal D}^\alpha_{abc}$
to the equations of motion (8) of  body $a$, caused by the
interaction of the other planets ($b\not=a, c\not=a,b$)
 with each other is:
{}
$${\cal D}^\alpha_{abc}={n^\alpha_{ab}\over
r_{ab}^2}\Big[(1-2\beta+2\zeta){1\over r_{bc}}-
2(\beta+\gamma-\zeta){1\over r_{ac}}\Big]-$$
{}
$$-(\zeta+\tau){\Pi^{\alpha\lambda}_{ab}
\over r^3_{ab}}({n_{bc}}_\lambda+{n_{ca}}_\lambda)
+(\zeta-\tau){{n_{ab}}_\lambda\over r_{ab}^2}
{\Lambda^{\alpha\lambda}_{ac}\over r_{ac}}+$$
{}
$$+{1\over 2}(1+2\zeta-2\tau){{n_{bc}}_\lambda\over r_{bc}^2}
{\Lambda^{\alpha\lambda}_{ac}\over r_{ac}}+
2(1+\gamma){{n_{bc}}^\alpha\over r_{bc}^2r_{ab}}, \eqno(14)$$

\noindent where the tensors
$\Lambda_{ab}^{\mu\nu}=\gamma^{\mu\nu}+n_{ab}^\mu n_{ab}^\nu$
and $\Pi_{ab}^{\mu\nu}=\gamma^{\mu\nu}+3n_{ab}^\mu n_{ab}^\nu$
are the projecting and polarizing operators respectively.

The metric tensor (1), the Lagrangian function (4) and the equations
of motion (8)  define  the
behavior  of the celestial  bodies in the post-Newtonian
approximation. Hence, they may be used for
producing the numerical codes in relativistic orbit determination
formalisms for planets and satellites
(Moyer, 1981; Ries {\it et al.}, 1991;
Standish {\it et al.}, 1992) as well as for
analysing   the gravitational experiments in the solar system
(Will, 1993; Pitjeva, 1993; Anderson {\it et al.}, 1996).

\subsection{Metric Tests of  Parametrized Gravitation.}

By means of a topographic Legendre expansion complete through the
second degree and order, the systematic error in Mercury radar
ranging has been reduced significantly (Anderson  {\it et al.}, 1995).
However, a {\it Mercury Orbiter} is required before significant
improvements in relativity tests become possible.  Currently, the
precession rate of Mercury's perihelion, in excess of the 530
arcsec per century ($''$/cy) from planetary perturbations, is
43.13 $''$/cy with a realistic standard error of 0.14  $''$/cy
 (Anderson  {\it et al.}, 1991).  After taking into account a small
excess precession from the solar oblateness, Anderson et al. find
that this result is consistent with general relativity.  Pitjeva
(1993) has obtained a similar result but with a smaller estimated
error of 0.052  $''$/cy.  Similarly, attempts to detect a
time variation in the gravitational constant $G$ using Mercury's
orbital motion have been unsuccessful, again consistent with
general relativity.  The current result (Pitjeva, 1993) is
$\dot{G}/G = (4.7 \pm 4.7) \times 10^{-12} \hbox{ yr}^{-1}$.

Metric tests utilizing a  {\it  Mercury
Orbiter}  have been studied both at JPL and at the Joint Institute
for Laboratory Astrophysics (JILA) and the University of
Colorado.  The JPL studies, conducted in the 1970's, assumed that
orbiter tracking could provide daily measurements (normal points)
between the Earth and Mercury centers of mass with a 10~m
standard error.  A covariance analysis was performed utilizing a
16-parameter model consisting of six orbital elements for Mercury
and Earth respectively, the metric relativity parameters $\beta$
and $\gamma$, the solar quadrupole moment $J_{2\odot}$, and the
conversion factor A between unit distance (Astronomical Unit) and
the distance in meters between Earth and Mercury.  It was assumed
that no other systematic effects were present, and that the
normal-point residuals after removal of the 16-parameter model
would be white and Gaussian.  The total data interval, assumed
equal to two years, corresponded to 730 measurements.  Under the
assumed random distribution of data, the error on the mean
Earth-Mercury distance was $10/\sqrt{730} = 37$ cm.  The JPL
studies showed, based on a covariance analysis, that the primary
metric relativity result from a {\it Mercury Orbiter} mission would be
the determination of the parameter $\gamma$, the parameter that
measures the amount of spatial curvature caused by solar
gravitation.  The standard error was 0.0006, about a factor of
two improvement over the Viking Lander determination.  This
accuracy reflected the effect of spatial curvature on the
propagation of the ranging signal and also its effect on
Mercury's orbit, in particular the precession of the perihelion.
The error in the metric parameter $\beta$ and the error in the
solar $J_{2\odot}$ were competitive with current results, but not
significantly better.

Within the last five years, a more detailed covariance analysis
by the JILA group (Ashby et al., 1995) assumed 6~cm ranging
accuracy over a data interval of two years, but with only 40
independent measurements of range.  Unmodeled systematic errors
were accounted for with a modified worst-case error analysis.
Even so, the JILA group concluded that a two-order of magnitude
improvement was possible in the perihelion advance, the
relativistic time delay, and a possible time variation in the
gravitational constant G as measured in atomic units.  However,
the particular orbit proposed by ESA for its 2000 Plus mission was
not analyzed.  It is almost certain that the potential of the ESA
mission lies somewhere between the rather pessimistic JPL error
analysis and the JILA analysis of an orbiter mission more nearly
optimized for relativity testing.

\subsubsection{Mercury's Perihelion Advance.}

The secular trend of Mercury's
perihelion in  the four-dimensional ({\small 4D})
 {\small PPN} formalism  depends
on the linear combination of the {\small PPN}
parameters $\gamma$ and $\beta$ and the solar quadrupole
coefficient $J_{2\odot}$ (Nobili and Will, 1986;
Heimberger {\it et al.}, 1990; Will, 1993):
{}
$$\dot{\pi}_{4D} =  (2+2\gamma-\beta)
{  GM_\odot n_M \over c^2a_M(1-e^2_M)} +
 {3\over 4}\Big({R_\odot \over a_M}\Big)^2
{J_{2\odot}n_M\over (1-e^2_M)^2 } (3\cos^2 i_M-1),
\hskip 10pt ''/\hbox{cy} \eqno(15a)$$
\noindent where $a_M, n_M, i_M$ and $e_M$ are the
mean distance, mean motion, inclination and eccentricity of Mercury's orbit.
The parameters $M_\odot$ and $R_\odot$ are the solar mass  and
radius respectievely.
For the  Mercury's orbital parameters  one obtains:
{}
$$\dot{\pi}_{4D} = 42 {''\hskip -3pt .}98 \hskip 2pt
\Big[ \hskip 1pt{1\over 3}(2+2\gamma-\beta) +0.296
\cdot {J_{2\odot}}\times 10^4\Big], \hskip 10pt ''/\hbox{cy}  \eqno(15b)$$

Thus, the accuracy of the relativity tests on the {\it Mercury Orbiter}
 mission will
depend on   our knowledge of the solar gravity field.
The major source of  uncertainty in these measurements is
the solar quadrupole  moment $ {J_2}_\odot$.
  As evidenced by the oblateness of the photosphere
(Brown  {\it et al.}, 1989) and perturbations in frequencies of solar
oscillations, the internal structure of the Sun is slightly
aspherical.  The amount of this asphericity is uncertain.
It has been suggested that it could be significantly larger than
calculated for a simply rotating star, and that the internal
rotation rate varies with the
solar cycle (Goode and Dziembowski 1991).  Solar oscillation data
suggest that most of the Sun rotates slightly slower than the
surface except possibly for a more rapidly rotating core (Duvall and
Harvey, 1984).  An independent measurement  of $ {J_2}_\odot$
performed  with the {\it Mercury Orbiter}
would provide a valuable direct confirmation of the indirect
helioseismology value  Furthermore, there are suggestions of a
rapidly rotating core, but helioseismology determinations have been
limited by uncertainties for depths below 0.4 solar radii
(Libbrecht and Woodard, 1991).

The {\it Mercury Orbiter} will help us understand  this
asphericity and  independently will enable us to gain some
important data on the properties of the solar interior and the
features of it's rotational motion.  Preliminary
analysis of a {\it Mercury Orbiter} mission suggests
that $J_{2\odot}$ can be determined to at best
 $\sim 10^{-9}$  (Ashby, Bender and Wahr, 1995).
Even a determination ten times worse, on the order of
10\% of the total effect, would be comparable in accuracy to the
solar oscillation determination (Brown  {\it et al.}, 1989)

In addition to this, the studies of Mercury's perihelion
advance might provide  us with an opportunity to  test  the hypothesis of
the multi-dimensionality ({\small D} $>4$) of our physical
world (Gross and Perry, 1983; Lim and Wesson, 1992;
Kalligas {\it et al.}, 1995). As it was
shown by Lim and Wesson, multi-dimensional extensions of
general relativity may produce  observable contributions
to the four-dimensional physical picture. It has been conjectured that
even a small extra (curved)
dimension may affect the dominant terms in the four-dimensional metric,
thereby altering the theoretical value of the perihelion shift.
For example, in the five-dimentional Kaluza-Klein theory, the
predicted value of  Mercury's perihelion advance is given as
{}
$$\dot{\pi}_{5D} = \dot{\pi}_{4D}(1+k^2), \eqno(15c)$$
\noindent where $k$ is the small  parameter
representing the motion of the body in the extra fifth dimension.

It should be noted that  the {\it Mercury Orbiter} itself,
being placed on orbit around Mercury,   will experience the
phenomenon of periapse advance as well.
However,
we expect that uncertainties in Mercury's gravity field will mask the
relativistic precession, at least at a level of interest for
ruling out alternative gravitational theories.

\subsubsection{The Precession Phenomena.}

In addition to the perihelion advance metric theories predict several
precession phenomena associated with the angular momentum of the
bodies.
Thus, according to {\small PPN} formalism,
the spin of a gyroscope  $\vec{s_0}$ attached to a test body orbiting
around the Sun precesses with respect to a distant standard of
rest - quasars or distant galaxies. The theory of  Fermi-Walker
transport describes the motion of the spin vector of a gyroscope
by the relation
{}
$${d  \vec{s_0}\over dt}=[ \vec{\Omega} \times  \vec{s_0}].\eqno(16)$$

\noindent where the angular velocity $\vec{\Omega}$ has the following
form (Will, 1993):
{}
$$\vec{\Omega}=-{1\over 2}[ \vec{v}\times \vec{a} ]+
( \gamma+{1\over 2})[ \vec{v}\times\vec{\nabla}U_\odot ]-$$
$$- (1+\gamma)
{\mu_\odot\over 2r^3 }\Big(\vec{S_\odot}-
{3(\vec{S_\odot}\vec{n})\vec{n}}\Big). \eqno(17)$$

\noindent Here $\vec{v}$  is the velocity of the body,
$\vec{a}=d\vec{v}/d\tau  -\vec{\nabla}U_\odot $
 is the non-gravitational
acceleration of the body under consideration
(for example from  the solar
radiation pressure),
$\vec{r} $ the body's heliocentric radius vector,
and $U_\odot$ is the gravitational potential of
the Sun at the body's location.

The first term in  (17) is the Thomas precession.
It is a special relativistic effect due to the non-comutativity of
nonaligned
Lorentz transformations (Thomas, 1927; Bini {\it et al.}, 1994).
It also can be interpreted as a coupling between
the bodies velocity and the non-gravitational forces acting on it.
The studies of this precession may provide an additional knowledge
of the non-gravitational environment of Mercury and its
interaction with the solar wind.
As for the quasi-Lorentz transformations to the proper coordinate
reference frame,
this term  produces a negligible small contribution to the  total
precession rate of the satellite's orbit.

The second term is known as a geodetic precession (De-Sitter, 1916).
This term arises in any non-homogeneous gravitational field  because of the
parallel transport of a direction defined by $\vec{s_0}$. It can be
viewed as spin precession caused by a  coupling
between the particle velocity $\vec{v}$ and the static
part of the space-time geometry (1).
For  Mercury orbiting the Sun this precession has the form:
{}
$$\vec{\Omega}_G= (\gamma +{1\over 2})
{\mu_\odot\over r^3}(  \vec{r}\times \vec{ v}), \eqno(18)$$
\noindent where $\mu_\odot$ is the solar gravitational constant,
$r$ is the distance from   Mercury to the Sun and $v$ is its
the orbital velocity.
This effect should be studied for the {\it Mercury Orbiter}, which, being
placed in orbit around  Mercury is in effect a gyroscope orbiting the Sun.
Thus, if we    introduce the angular momentum per unit mass,
$ \vec{ L}= \vec{ r}\times  \vec{ v}$,
of   Mercury in solar orbit, the equation (18) shows
that $ \vec{ \Omega}_G$ is directed along the pole of ecliptic,
in the direction of $ \vec{ L}$.
The vector $ \vec{\Omega}_G$ has constant part
{}
$$ \vec{\Omega}_0={1\over 2}(1+2\gamma)
 {\mu_\odot\omega_M\over a_M}=  {1+2\gamma\over 3}\cdot 0.205
 \hskip 10pt  { ''/\hbox{yr}}, \eqno(19a) $$

\noindent with the significant correction
  due to the eccentricity $e_M$
of the Mercury's orbit,
{}
$$ \vec{\Omega}_1 \cos \omega_M t=
 {3\over 2}(1+2\gamma) {\mu_\odot\omega_M\over a_M} e_M\cos \omega_M t=
{1+2\gamma\over 3}\cdot0.126 \cos \omega_M t
 \hskip 10pt  {''/\hbox{yr}}, \eqno(19b) $$

\noindent where $\omega_M$ is   Mercury's siderial frequency and
$t$ is reckoned from a perihelion passage;
$a_M$ is the semimajor axis of  Mercury's orbit.

This effect has been studied for the motion of lunar perigee
and the existence of the geodetic precession was confirmed with
an accuracy of 10\% (Bertotti, Ciufolini and Bender, 1987).
Two other groups has analyzed the lunar laser-ranging data
to estimate the deviation of the lunar orbit from the
predictions of general relativity
 (Shapiro {\it et al.}, 1988; Dickey {\it et al.}, 1989).
  These predictions
have been confirmed  within the standard deviation of 2\%.

Certainly this experiment, being performed in the
vicinity of the Mercury, requires
complete analysis, taking into account the
real orbital parameters of the spacecraft.
Our  preliminary analysis
shows that the precession of the satellites's
orbital plane should  include a contribution of
order 0.205 ${''/\hbox{yr}}$  from the geodetic precession.
Therefore   special studies of this
precession  should be included in future studies of the
{\it Mercury Orbiter} mission.

The third term in the expression (17) is known as
Lense-Thirring precession $\vec{\Omega}_{LT}$.
This term gives the
relativistic precession of the gyroscope's spin
$\vec{s}_0$ caused by the intrinsic angular momentum
$\vec{S}$ of the central body.
This effect responsible for a small perturbation of orbits
of artificial satellites around the Earth (Tapley  {\it et al.},
1972;  Ries  {\it et al.}, 1991).
Unfortunately,  preliminary studies have shown that
this effect is also negligibly small
for the satellite's orbit around Mercury
and, moreover, it is
masked by the uncertainties of the orbit's inclination.

\subsubsection{The Test of the Time
Dependence of the Gravitational Constant.}

As pointed out by Dirac, the age of the
universe, $H_0^{-1}$, in terms
of the atomic unit of time, $e^2/{mc^3}$,
is of order $10^{39}$, as is the ratio,
$e^2/(Gm_pm)$, of the electrical force between the
electron and proton to the gravitational
force between the same particles.
This, according to Dirac, suggested that both ratios are
functions of the age of the universe, and
that the gravitational constant, $G$, might vary with time.
In order to  account for this possibility, and assuming that
$\dot{G}=dG/dt$ is constant in first approximation,  we define a
parameter $\lambda_G$ as follows: $\dot{G}/G=-\lambda_G H_0$.
Then, the effect of $\lambda_G$ on the motion of
bodies in the solar system can be
introduced as an additional perturbation in the
barycentric equations of motion
(8)   in the following form:
{}
$$\delta_{\lambda_G} \ddot{\vec{r}}=\lambda_G H_0
\sum_{b\not=a}{m_a t \over r_{ab}^3} \vec{n}_{ab} +{\cal O}(t^2).
\eqno(20)$$

\noindent The major contribution to this effect for the
motion of  celestial bodies in the solar system is the Sun.
The importance of the {\it Mercury Orbiter}
 mission will be in providing
the tracking data necessary for establishing a more accurate
determination of the  orbital elements
of this planet. Then, combined together
with other solar system data (Chandler {\it et al.}, 1994;
Anderson {\it et al.}, 1996), this information will enable one
to perform a more accurate test of the hypothesis (20).

\subsubsection{The Planetary Test of the Equivalence Principle.}

The development of the parameterized post-Newtonian
(PPN) formalism, has provided  a useful framework for testing
the violation of the Strong Equivalence Principle (SEP) for
extended bodies (Anderson  {\it et al.}, 1996). In that formalism,
the ratio of
passive gravitational mass ${m_b}_{(g)}$ to inertial
mass ${m_b}_{(i)}$ is given by
{}
$$ {{m_b}_{(g)} \over {m_b}_{(i)}} = \hskip2pt 1 \hskip2pt +
\hskip2pt \eta  {\Omega_b \over m_b c^{2}},
\hskip10pt  \eqno(21) $$

\noindent
where $m_b$ is the rest mass of a body $b$ and
$\Omega_b=\gamma_{\mu\nu}\Omega^{\mu\nu}_b$ is
its gravitational binding
energy (2.7). Numerical evaluation of the integral
of expression (7) for the standard solar model
 (Ulrich,  1982) gives
$$ \left({\Omega \over mc^{2}}\right)_S  \approx -3.52
\hskip1pt \cdot 10^{-6}   \eqno(22)$$

The  {\small SEP} violation  is quantified in eq.(21) by
the parameter $\eta$. In fully-conservative,
Lorentz-invariant theories of gravity
the  {\small SEP} parameter is related to
 the {\small  PPN} parameters by
{}
$$ \eta = 4 \beta - \gamma - 3  \hskip10pt  \eqno(23a) $$

\noindent and is more generally related to the complete set of
{\small  PPN} parameters through the relation (Will, 1993)
{}
$$ \eta = 4 \beta - \gamma - 3 -
{10 \over 3}\hskip 1mm\xi - \alpha_1 +
{1\over 3}\hskip 1mm (2 \alpha _2 -
2 \zeta_1 - \zeta_2). \hskip10pt
\eqno(23b) $$

A difference between gravitational and inertial
masses produces observable  perturbations in the
motion of celestial bodies in the solar system. By analyzing the
effect of a non-zero $\eta$
on the dynamics of the Earth-Moon system moving in the
gravitational field of the Sun, Nordtvedt (1968b) found  a
polarization of the Moon's orbit in the direction of the Sun
with amplitude $\delta r \sim \eta C_0$, where $C_0$ is a
constant of order $13m$ (Nordtvedt effect).
The most accurate test of this effect is presently
provided by Lunar Laser Ranging (LLR)
(Shapiro  {\it et al.}, 1976; Dickey  {\it et al.}, 1989),
and in the most  recent  results (Dickey  {\it et al.}, 1994;
Williams   et al. 1995) the parameter $\eta$ was determined to be
{}
$$\eta = -0.0005 \pm 0.0011 \hskip2pt.   \eqno(24)$$

\noindent
Also results are available from numerical experiments  with
combined processing of  LLR, spacecraft tracking,
planetary radar and Very Long Baseline Interferometer (VLBI) data
(Chandler  et al. 1994).
Recently, the analysis of the accuracy for
 planetary SEP violation in Sun-Jupiter-Mars-Earth system has
been done for future Mars missions (Anderson et al., 1996).
We would like  to emphasize that   a measurement of the sun's
gravitational to inertial mass ratio can be
performed using the  Mercury ranging experiment.
Indeed, let us consider, for example,  the dynamics of the
four-body Sun-Mercury-Earth-Jupiter
system in the solar system barycentric inertial reference frame.
The quasi-Newtonian acceleration of  Mercury $(M)$ with respect
to the Sun $(S)$,  $\ddot{\vec{r}}_{SM}$, is straightforwardly
calculated  from the equations (8) to be:
{}
$$ \ddot{\vec{r}}_{SM}=\ddot{\vec{r}}_M - \ddot{\vec{r}}_S =
-m^*_{SM} \cdot{\vec{r}_{SM} \over r_{SM}^3}
+ m_J \Bigl [ { \vec{r}_{JS} \over r_{JS}^3} - {\vec{r}_{JM} \over r_
{JE}^3}  \Bigl ] + \eta  \Bigl ( {\Omega \over mc^2} \Bigl )_{\hskip-2pt S}
m_J {\vec{r}_{JS} \over r_{JS}^3},  \eqno(25) $$

\noindent where  $m^*_{SM} \equiv m_S + m_M + \eta\Big[m_S
\Big( {\Omega \over mc^{2}} \Big )_{\hskip-2pt M} +
 m_E \Big({\Omega \over mc^2}  \Big)_{\hskip-2pt S}\Big ]$.
The first and  second terms on the right side of the
equation (25) are the Newtonian and the
 tidal acceleration terms respectively.
We will denote the last term in this equation as  $\vec{A}_{\eta}$.
This is the {\small SEP} acceleration term which is of order
$c^{-2}$ and  it is treated as a perturbation on the
restricted three-body problem.  While it is not the only term of
that order,  the other post-Newtonian $c^{-2}$ terms
(suppressed in eq.(25)) do not
affect the determination of $\eta$
until the second post-Newtonian order ($\sim c^{-4}$).
The corresponding  {\small SEP}
 effect is evaluated as an alteration of the planetary Keplerian orbit.
The subscripts $(E)$ and $(J)$ indicate
Earth and Jupiter, respectively. To good approximation the
{\small SEP} acceleration $\vec{A}_{\eta}$
has constant magnitude and  points in the direction from Jupiter
to the Sun, and because it  depends only on the mass
distribution in the Sun, the
Earth and Mercury experience the same perturbing acceleration.
The responses of
the trajectories of each of these planets  due to the   term
$\vec{A}_{\eta}$ determines the perturbation in the
 Earth-Mercury range and  allows
a detection of the  {\small SEP} parameter
$\eta$ through a ranging experiment.

The presence of the  acceleration term $\vec{A}_{\eta}$ in the
equations of motion  (25) results in  a
polarization of the orbits of Earth and Mercury,  exemplifying the
planetary {\small SEP} effect.
We have  examined this effect on the orbit of Mercury by carrying out
first-order perturbation theory about the
zeroth order of  Mercury's orbit, taken to be circular.
Then the  {\small SEP} perturbation produces
the polarization   of the Mercury's orbit  of order:
$\delta r = \eta  C_M$, where $C_M$  is a constant of order 81 m.
 However,  it turns out that  the eccentricity
correction  played a   significant role  in the similar problem of
Mars motion (Anderson et al., 1996).
 One reason for this is that these
 corrections  include  ``secular'' matrix elements which are
 proportional to the time $t$.
Such  elements  dominate at large times, and
  the eccentricity corrections thereby qualitatively change
the nature of the solution in the linear approximation.
Thus,   for a given value of  {\small SEP} parameter $\eta$,
 the polarization effects on the Mercury orbit are
more than two orders of magnitude larger than on the lunar orbit.
The  additional  result of these studies is that the mass
of Jupiter, $m_J$, can be determined more
accurately from a few years of Earth-Mercury ranging than from
Pioneer 10,11 and Voyager 1,2 combined. This analysis
shows a rich opportunity for  obtaining new scientific
results from the  program of ranging measurements to
Mercury during  the {\it Mercury Orbiter} mission.
Certainly, for the future analysis of the general planetary
{\small SEP} violation problem one
should modify the theoretical model to include effects due
to  Venus, Mars and  Saturn and  perform
 numerical experiments with combined data collected
from the planetary missions, {\small LLR} and {\small VLBI}.

Moreover, the analysis of {\it Mercury Orbiter}  ranging data
might provide the opportunity for another fundamental test, namely a
solar system search for dark matter.
The data obtained during this mission will allow one to study the
conditions leading to the violation of the strong equivalence
principle  in the motion of
 extended bodies in the solar system resulting from
 possible  composition, shape and rotation dependent
coupling of   dark
matter to the matter of  different astrophysical objects.

\subsubsection{The Redshift Experiment.}

Another important experiment that could be performed on a {\it Mercury
Orbiter} mission is a test of the solar gravitational redshift,
provided a stable frequency standard is flown on the spacecraft.
The experiment would provide a fundamental test of the theory of
general relativity and the Equivalence Principle upon which it
and other metric theories of gravity are based (Will, 1993).
Because in general relativity the gravitational redshift of an
oscillator or clock depends upon its proximity to a massive body
(or more precisely the size of the Newtonian potential at its
location), a frequency standard at the location of Mercury would
experience a large, measurable redshift due to the Sun.
 Moreover, the eccentricity of Mercury's orbit would be highly effective in
varying the solar potential at the clock, thereby producing a
distinguishing signature in the redshift. The anticipated frequency
variation between perihelion and aphelion is to first-order in
the eccentricity:
{}
$$\Big({\delta f\over f_0}\Big)_{e_M}=
{2\mu_\odot e_M \over  a_M}.\eqno(26)$$

\noindent This contribution
is quite considerable and is calculated to be
$(\delta f/f_0)_{e_M}=1.1\times 10^{-8}$.
It's absolute magnitude, for instance,
at the wave length $\lambda_0=3$ cm ($f_0=10$ GHz)
is $ (\delta f)_{e_M}=  110$ Hz.
 We would also benefit
from the short orbital period of Mercury, which would permit the
redshift signature of the Sun to be measured several times over
the duration of the mission. Depending upon the stability of the
frequency standard, a {\it Mercury Orbiter} redshift experiment could
provide a substantial improvement over previous tests of the
solar redshift.

At this time, there exist two comparable tests of the solar
redshift. The first test was performed by using the well-known
technique of measuring shifts in the positions of spectral lines
of atoms in the solar atmosphere. Recent measurements of an
oxygen triplet have verified the solar redshift predicted by
general relativity to an accuracy of $1\%$ (LoPresto {\it et al.},
1991). The first test of the solar redshift with an
interplanetary probe was performed recently with the {\it Galileo}
spacecraft during its mission to Jupiter (Krisher, 1993). In
order to obtain gravity-assists from the Earth and Venus, the
spacecraft traveled on a trajectory that took it in and out of
the solar potential. By measuring the frequency of an
ultra-stable crystal oscillator (USO) flown on the {\it Galileo}
spacecraft during this phase of the mission, the redshift
prediction was again verified to an accuracy of $1\%$. The
accuracy obtained was limited by the long-term stability of the
USO. After calibration of systematic errors (e.g., a linear drift
due to aging), the fractional frequency stability was roughly
$10^{-12}$.

During a {\it Mercury Orbiter} mission, the variation in the solar
potential between perihelion and aphelion would produce a
redshift of roughly 1 part in $10^8$ according to general relativity
(for a detailed analysis of the anticipated redshift,
see Krisher, 1993). With a fractional frequency stability of
$10^{-12}$, we could test the redshift with an accuracy of 1 part
in $10^4$ (or $0.01\%$). This level of accuracy is comparable to
that obtained by the 1976 Vessot/NASA Gravity-Probe A (GP-A)
experiment performed in the gravitational field of the Earth with
a hydrogen maser on a Scout rocket (Vessot et al., 1980). Greater
sensitivity could be obtained if an atomic frequency standard
were flown on {\it Mercury Orbiter}. An atomic standard having a
typical stability of 1 part in $10^{15}$ would permit the redshift to be
tested to 1 part in $10^7$.

\subsubsection{The Shapiro Time Delay.}

According to general relativity, an electromagnetic signal propagating
near the Sun will be delayed by solar gravity (Shapiro, 1964).
In {\small PPN} formalism this additional
coordinate time delay, $\Delta \tau $,
is given by (Will, 1993; Lebach {\it et al.}, 1995)
{}
$$c\Delta \tau = \mu_\odot\Big[(1+\gamma)
\ln\Big({r_E+r_p+r_{Ep}\over r_E+r_p-r_{Ep}}\Big)-$$
$$-{\tau\over 2r_{Ep}}\Big({r^2_E -r^2_p-r_{Ep}^2\over r_p}+
{r^2_p-r^2_E-r_{Ep}^2\over r_E}\Big)\Big] \eqno(27) $$

\noindent where $r_E$ and $r_p$ are the respective
distances of the Earth and the planet from the Sun,
and $\vec{r}_{Ep}=\vec{r}_p-\vec{r}_E$ is the earth-planet distance.
At superior conjunction this formula can be approximated by
{}
$$c\Delta \tau =\mu_\odot\Big[(1+\gamma)\ln\Big({
4r_Er_p\over d^2}\Big)+2\tau\Big] \eqno(28) $$
\noindent where $d$ is the closest distance from
the center of the Sun to the beam trajectory.

By ranging to {\it Mariner}, {\it Viking}, and {\it Voyager}
 spacecraft during
solar conjunction, it has been possible to test
this effect to a few percent or better (Krisher {\it et al.}, 1991 and
references therein).  From maximum elongation to superior conjunction,
$\Delta\tau$ increases from 15 to 240 microseconds for Mercury.
However, all previous spacecraft tests have depended on an S-band uplink.
The {\it Mercury Orbiter} may provide the first
opportunity for an X-band test. The higher frequency, by the factor
11/3, will reduce  the systematic error from propagation of the radio beam
through the solar corona. It will also be possible to perform a Doppler
relativity test as suggested by Bertotti and Giampieri (1992).
The multi conjunctions may be useful in constraining certain
alternative theories of gravitation involving anisotropic metric potentials.
The Doppler shift caused by solar gravity could depend on
the geometry of the conjunction.

\section{Radio Science Instrumentation}

The {\it Mercury Orbiter} radio science objectives require on
board instrumentation with the following capabilities:

1. Up-link  at X-band ($\sim 3.5$ cm wavelength)
           from ground stations to the spacecraft.

2. X-band down-link coherently referenced to   the  up-link signal.

3. Crystal or atomic ultra-stable oscillator (USO).

4. Simultaneous down-link at  S ($\sim 13$ cm) and X-bands
           referenced to the USO.

Any S-band up-link capability, perhaps provided for commanding
or emergency communications, would  not be used during radio science
observations.
The  S-band   up-link provides far superior performance.
Up-link and down-link at higher frequencies
(i.g. KA-band) should be considered for the {\it Mercury Orbiter}
mission for purposes of enhancing the  gravity
and relativity measurements.
A multi-link Doppler systems should be considered as well.
Oven controlled USO's has been included in several radio science
systems in the past Including
{\it Voyager}, {\it Galileo} and {\it Cassini}.
The {\it Mercury Orbiter} mission is far enough in the future
that the crystal oscillator will be replaced by an atomic standard.

Starting with an X-band system, the cost of adding a
dual-frequency downlink capability to enhance radio propagation
measurements would require about 15-20 watts of additional power
and 4-5 kg of additional mass for an S-band transmitter.
Observations could then be performed at both occultation ingress
and egress by transmitting both frequencies coherently from a
common on-board oscillator.    It would not be necessary to add a
special ultra-stable oscillator (USO) for this purpose as long as
the downlinks are phase coherent.

\section{Acknowledgements}
The research described in this paper was carried out by the Jet
Propulsion Laboratory, California Institute of Technology, and
was sponsored by the Ultraviolet, Visible, and Gravitational
Astrophysics Research and Analysis Program through an agreement
with the National Aeronautics and Space Administration.

\section{References}
\begin{itemize}
\item[] {\bf Anderson, J. D.,  Colombo, G.,  Esposito, P. B.,  Lau, E. L.
        and Trager, G. B.,} ``The mass, gravity field, and ephemeris of
        Mercury''. \it Icarus, \bf 71\rm,  337-349, 1987.
\item[] {\bf Anderson, J. D., Slade,  M. A.,  Jurgens, R. F.,  Lau, E. L.,
        Newhall, X X and Standish, E. M., Jr.,} ``Radar and spacecraft
        ranging to Mercury between 1966 and 1988''. \it Proc. Astron. Soc.
        Australia, \bf 9\rm, 324, 1991.
\item[] {\bf Anderson, J. D., Jurgens, R. F., Lau, E. L.,   Slade, M. A.,
        III, and Schubert, G.,} ``Shape and orientation of Mercury
        from radar ranging data''. Preprint, 1995.
\item[] {\bf Anderson, J. D.,   Gross, M.,  Nordtvedt  K.,
        and Turyshev,  S. G.,} ``The Solar Test of the
        Equivalence Principle''.
        To appear in March 1 of the {\it ApJ},   1996.
\item[] {\bf Ashby, N.,  Bender, P. L. and Wahr, J. M.,} ``Gravitational
        physics tests from ranging to a Mercury transponder satellite''.
        Preprint, 1995.
\item[] {\bf Balmino, G., Lambeck,  K. and  Kaula, W. M.,} ``A spherical
        harmonic analysis of the Earth's topography''. \it J. Geophys.
        Res., \bf 78\rm, 478 - 481, 1973.
\item[] {\bf Bertotti, B., Ciufolini, I.,  and Bender, P.,} ``New test of
        General Relativity: Measurement of de Sitter
        Geodetic Precession Rate
        for Lunar Perigee''.
       {\it Phys. Rev. D}, {\bf58},  1062-1065,   1987.
\item[] {\bf Bertotti, B. and  Giampieri, G.,}
       ``Relativistic effects for Doppler
        measurements near solar conjunction''.
        {\it Class. Quantum. Gravity}. {\bf 9}, 777-793, 1992.
\item[] {\bf Bills, B. G., and  Ferrari, A. J.,} ``A harmonic analysis of
        lunar topography''. \it Icarus, \bf 31\rm, 244 - 259, 1977.
\item[] {\bf Bills, B. G., and Ferrari,}  A. J.,  ``Mars topography
        harmonics and geophysical implications''. \it J. Geophys.
        Res., \bf83\rm, 3497 - 3508, 1978.
\item[] {\bf Bindschadler, D. L., Schubert, G. and  Ford, P. G.,} ``Venus'
        center of figure - center of mass offset''. \it Icarus,
        \bf 111 \rm, 417 - 432, 1994.
\item[] {\bf Bini, D., Carini, P., Jantzen, R. T., Wilkins, D.,}
        ``Thomas precession in post-Newtoni- an gravitomagnetism''.
        {\it Phys. Rev. D.}, {\bf49}(6), 2820-27, 1994.
\item[] {\bf Bird, M.K., Asmar,  S.W.,  Brenkle, J. P.,  Edenhofer, P.,
         P\"{a}tzold, M. and  Volland, H.,} ``The coronal-sounding
        experiment''. {\it Astron.\ Astrophys.\ Supp.\ Ser.\ }{\bf 92},
        425-430, 1992.
\item[] {\bf Brown, T.M., Christensen-Dalsgaard, J.,  Dziembowski, W.A.,
        Goode,  P.,  Gough,  D.O. and  Morrow, C.A.,} ``Inferring the Sun's
        Internal Angular Velocity from Observed p-Mode Frequency Splittings''.
        {\it ApJ }, {\bf343}, 526-546,  1989.
\item[] {\bf Brumberg, V. A.,} {\it Essential
        Relativistic Celestial Mechanics}.
        Hilger, Bristol,  1991.
\item[] {\bf Chandler, J. F.,  Reasenberg, R. D. and  Shapiro I. I.,}
        ``New Results on the Principle of Equivalence''.
        {\it BAAS}, {\bf 26},  1019,  1994.
\item[] {\bf Chandrasekhar, S., F.R.S., and  Contopulos, G.,}
        ``On a post-Galilean transformation appropriate to the
        post-Newtonian theory of Einstein, Infeld and Hoffmann''.
       {\it Proc. Roy. Soc. London}, {\bf A298},  123-141,  1967.
\item[] {\bf Cheng, A.F., Johnson,  R.E., Krimigis, S.M. and  Lanzerotti,
L.J.,}
        ``Magnetosphere, exosphere and surface of Mercury''.
        {\it Icarus}, {\bf 71}, 430-440, 1987.
\item[] {\bf Damour, T.,} ``The problem of motion in Newtonian and
        Einsteinian gravity''.
        In: {\it Three Hundred Years of Gravitation},
        eds. S. W. Wawking and W. Israel,
        Cambridge University Press,   128-198,   1983.
\item[] {\bf Denisov, V. I., and  Turyshev, S. G.,}
        ``Motion of an extended bodies in the parametrized
        post-Newtonian formalism''.
        {\it Theor. Mat. Fiz.}, {\bf 83}(1),   129-135,   1990.
\item[] {\bf De-Sitter, W.,} ``On Einstein's Theory of Gravitation,
        and its Astronomical Consequences. First Paper''.
        {\it Mon. Not. Roy. Astron. Soc.}
        {\bf 76},  699-728, 1916.
\item[] {\bf Dickey, J. O., Newhall, X. X. and Williams, J. G.,}
        ``Investigating Relativity using Lunar Laser Ranging:
         Geodetic Precession and the Nordtvedt Effect''.
        {\it Advance in Space Research}, {\bf 9}(9), 75-78, 1989.
\item[] {\bf Dickey, J. O. {\it and 11 co-authors},}
        ``Lunar Laser Ranging: A Continuing Legacy of the
        Apollo Program''.
        {\it Science}, {\bf 265},  482-490,  1994.
\item[] {\bf Duvall, T.L., Jr. and  Harvey, J.W.,}  ``Rotational Frequency
        Splitting of Solar Oscillations''.  Nature, {\bf 310},  19-22,  1984.
\item[] {\bf ESA Study Team,} {\it M3 selection process: Presentation of
        assessment study results},   ESA Doc.  No. SCI({\bf94}) 9,
        May 1994.
\item[] {\bf Fock,  V. A.,}  {\it  The theory of Space, Time and Gravitation}.
        Pergamon, Oxford, 1959.
\item[] {\bf Fjeldbo, G., Kliore, A.,  Sweetnam, D., Esposito,  P.,
        Seidel, B., and   Howard, T.,} ``The occultation of Mariner 10 by
        Mercury'', \it Icarus, \bf 29\rm, 439-444, 1976.
\item[] {\bf Fukushima, T.,} ``The ephemeris''. {\it A}\&{\it A}, {\bf 294},
        895-906, 1995.
\item[] {\bf Goode, P.R. and Dziembowski, W.A.,}
        ``Solar-Cycle Dependence of the Sun's Deep Internal Rotation
        Shown by Helioseismology''.  {\it Nature}, {\bf349},  223-225,  1991.
\item[] {\bf Gross, D. J., and Perry, M. J.,}
        ``Magnetic monopoles in Kaluza-Klein theory''.
       {\it Nucl. Phys.} {\bf B, 226}(1), 29-48, 1983.
\item[] {\bf Harmon, J. K.,  Campbell, D. B., Bindschadler,  D. L.,
        Head, J. W., and  Shapiro, I. I.,} ``Radar altimetry of Mercury: A
        preliminary analysis''. \it J. Geophys. Res., \bf 91\rm,
        385-401, 1986.
\item[] {\bf Harmon, J. K, and  Campbell, D. B.,} ``Radar observations of
        Mercury''. In \it Mercury, \rm F. Vilas, C. R. Chapman,
        and M. S. Matthews, eds.,   The University of
        Arizona Press, Tucson,   101 - 117, 1988.
\item[] {\bf Hechler, M.,} {\it Mercury Orbiter} Assessment study mission
        analysis, {\it ESOC Doc.\ No. MERC-MA-WP-345-OD}, March 1994.
\item[] {\bf Heimberger, J., Soffel, M. H., and Ruder, H.,}
        ``Relativistic effects in the motion of
        artificial satellites: the oblateness of the central body II''.
        {\it Celestial Mechanics and Dynamical Astronomy},
        {\bf 47}, 205-217, 1990.
\item[] {\bf Howard, H.T., {\it and 20 co-authors},} ``Mercury: Results
        on mass, radius, ionosphere, and atmosphere from
        {\it Mariner~10} dual-frequency radio signals''. {\it Science},
        {\bf 185}, 179-180, 1974.
\item[] {\bf Ingalls, R. P. and  Rainville, L. P.,}  ``Radar measurements of
        Mercury: Topography and scattering characteristics at
        3.8 cm'', \it AJ, \bf 77\rm, 185-190, 1972.
\item[] {\bf Kalligas, D., Wesson, P.S., and Everitt, C. W. F.,}
        ``The classical tests in Kaluza-Klein gravity''.
        {\it ApJ}, {\bf 439}, 548-557, 1995.
\item[] {\bf Kaula, W. M.,  Schubert, G., Lingenfelter, R. E.,
        Sjogren,  W. L. and  Wollenhaupt, W. R.,} ``Analysis and interpretation
        of lunar laser altimetry''. \it Proc. Third Lunar Planet.
        Sci. Conf. \rm, 2189 - 2204, 1972.
\item[] {\bf Konopliv A. S., and   Sjogren, W. L.,} ``Venus Sherical Harmonic
        Gravity Model to Degree and Order 60''. {Icarus}, {\bf 112},
        42-54,  1994.
\item[] {\bf Kopejkin, S. M.,} ``Celestial Coordinate Reference Systems in
        Curved Space-Time''. {\it Celest. Mech.}, {\bf 44},  87, 1988.
\item[] {\bf Kliore, A.J., Fjeldbo,  G.,  Seidel, B.L.,
        \hyphenation{Swe-et-nam}, D.N.,
        \hyphenation{Ses-plau-kis}, T.T., and  Woiceshyn, P.M.,}
        ``The atmosphere of Io from
        {\it Pioneer~10} radio occultation measurements''. {\it Icarus},
        {\bf 24}, 407-410, 1975.
\item[] {\bf Krisher, T. P., Anderson,  J. D.,  and Taylor,  A. H.,}
        ``Voyager 2 test  of the radar time-delay effect''.
        {\it ApJ}, {\bf 373}, 665-670, 1991.
\item[] {\bf Krisher, T. P., Morabito,  D. D., and  Anderson, J. D.,}
        ``The Galileo Solar Redshift Experiment''.
        {\it Phys. Rev. Lett.}, {\bf 70}, 2213-2216, 1993.
\item[] {\bf Krisher, T. P.,} ``Parametrized Post-Newtonian
        Gravitational Redshift''.
       {\it Phys. Rev. D}, {\bf 48}, 4639-4644, 1993.
\item[] {\bf Lebach, D. E.,  Corey, B. E.,  Shapiro, I. I.,  Rathner,  M. I.,
        Webber, J. C.,  Rogegrs,  A. E. E.,  Davis, J. C. and
        Herring, T. A.,} ``Measurement of the Solar Gravitational
        Deflection of Radio Waves Using Very-Long-Baseline
        Inerferometry''. {\it Phys. Rev. Lett.}, {\bf 75}(8), 1439-1442,  1995.
\item[] {\bf Libbrecht, K.G. and  Woodard, M.F.,}
        ``Advances in Helioseismology''.  {\it Science}, {\bf253}, 152-157,
1991.
\item[] {\bf Lim, P. H., and Wessen, P. S.,} ``The perihelion problem in
        Kaluza-Klein gravity''. {\it ApJ}, {\bf 397}, L91-L94, 1992.
\item[] {\bf Lindal, G.F., Wood,  G.E., Hotz, H.B.,  Sweetnam, D.N.,
        Eshleman, V.R.
        and Tyler,  G.L.,} The atmosphere of Titan: An analysis of the
        {\it Voyager 1} radio occultation measurements. {\it Icarus},
        {\bf 53}, 348-363, 1983.
\item[] {\bf LoPresto, J. C., Schrader,  C., and  Pierce, A. K.,}
        ``Solar Gravitational Redshift from the Infrared Oxigen Triplet''
        {\it ApJ}, {\bf 376}, 757-760, 1991.
\item[] {\bf Moyer, T. D.,} ``Mathematical Formulation of the
        Double Precision Orbit Determination Program
        (DPODP)'', Jet Propulsion Laboratory  Technical Report 32-1527,
        Pasadena, California,   1971.
\item[] {\bf Moyer, T. D.,} ``Transfromation from Proper Time on
        Earth to Coordinate Time in Solar System Barycentric
        Space-Time Frame of Reference. Parts 1 and 2''.{\it Celest. Mech.},
       {\bf 23}, pp.33-57 and pp.57-68, 1981.
\item[] {\bf Newhall, X X., Standish, E. M. Jr., and Williams, J. G.,}
        ``DE-102 - A Numerically Integrated Ephemeris of the Moonand Planets
        Spanning 44 Centuries''.
        {\it A}\&{\it A}, {\bf 125}(1),  150-167, 1983.
\item[] {\bf Ni, W.-T.,  and Zimmermann, M.,}
        ``Inertial and Gravitational Effects in the Proper Reference
         Frame of an Accelerated, Rotating Observer''.
        {\it  Phys. Rev.} {\bf D}, {\bf17}, p.1473,  1978.
\item[] {\bf Nobili, A., and Will,  C. M.,}  The real value of Mercury's
        perihelion advance. {\it Nature}, {\bf 320}, 39-41,  1986.
\item[] {\bf Nordtvedt, K., Jr.,}
        ``Equivalence Principle for massive bodies. Parts 1 and 2''.
        {\it Phys. Rev.}, {\bf169}, pp.1014-16 and 1017-25, 1968a.
\item[] {\bf Nordtvedt, K., Jr.,}
        ``Testing relativity with laser ranging to the Moon''.
        {\it Phys. Rev.}, {\bf170}, 1186-7, 1968b.
\item[] {\bf Pitjeva, E. V.,} ``Experimental testing of relativistic
        effects, variability of the gravitational constant, and
        topography of Mercury surface from radar observations
        1964-1989''. \it Celestial Mechanics and Dynamical Astronomy,
        \bf 55\rm, 313-321, 1993.
\item[] {\bf Ries, J. C., Huang, C., Watkins, M. M., and Tapley, B. D.,}
        ``Orbit Determination in the Relativistic
         Geocentric Reference Frame''.
        {\it The Journ. of the Astronautical Sciences},
        {\bf 39}(2), 173-181, 1991.
\item[] {\bf Schubert, G.,  Anderson, J. D., Jurgens,  R. F.,  Lau,  E. L.,
        and Slade, M. A. III,} ``Shape and orientation of Mercury from
        radar ranging data''. AGU abstract, Preprint, 1995.
\item[] {\bf Shapiro, I. I.,} Fourth test of general relativity.
        {\it Phys. Rev. Lett.}, {\bf 13}, 789-791, 1964.
\item[] {\bf Shapiro, I. I.,  Zisk, S. H.,  Rogers,  A. E. E.,  Slade, M. A.
        and  Thompson, T. W.,} ``Lunar topography:  Global determination
        by radar''. \it Science, \bf 178\rm, 939-948, 1972.
\item[] {\bf Shapiro, I. I., Councelman, C. C., III, and King, R. W.,}
        ``Verification of the principle of equivalence  for massive bodies''.
        {\it Phys. Rev. Lett.}, {\bf 36}, 555-8, 1976.
\item[] {\bf Shapiro, I. I., Reasenberg, R. D.,
        Chandler, J. F., Babcock, R. W.,}
        ``Measurement of the de-Sitter Precession of the Moon -
        A Relativistic Three-Body Effect''.
        {\it Phys. Rev. Lett.}, {\bf 61}(23), 2643-2646, 1988.
\item[] {\bf Smith, D. E.,  Zuber, M. T.,  Neumann,  G. A., and
        Lemoine, F. G.,} The topography of the Moon from Clementine LIDAR.
        Submitted to \it J. Geophys. Res., \rm 1995.
\item[] {\bf Standish, E. M.,} ``The figure of Mars and its effect on
        radar ranging''. {\it A}\&{\it A}, \bf 26\rm, 463-466, 1973.
\item[] {\bf Standish, E. M.,} ``The \hyphenation{ob-ser-va-tional ba-sis}
        for JPL's DE200,
        the planetary ephemerides of the Astronomical Almanac''.
        {\it A}\&{\it A}, \bf 233\rm, 252-271, 1990.
\item[] {\bf Standish, E. M., Newhall, X X ,  Williams, J. G., and
        Yeomans, D. K.,} ``Orbital ephemerides of the Sun, Moon, and planets''.
        In: \it Explanatory Supplement to the Astronomical Almanac,
        \rm  ed. P. K. Seidelmann, pp. 279-323, Mill Valley:
        University Science Books, 1992.
\item[] {\bf Tapley, B. D.} ``Statistical Orbit Determination Theory'',
        In {\it NATO  Advanced Study Institute in Dynamical Astronomy,}
        ed. B. D. Tapley and V. Szebehely. D. Reidel, 396-425, 1972.
\item[] {\bf Thomas, L. H.,} {\it Philos. Mag.}, {\bf 3},  1,   1927.
\item[] {\bf Thorne, K. S.,}
        ``Multipole expansions of gravitational radiation''.
         {\it Rev. Mod. Phys.}, {\bf 52}(2),  299-339,  1980.
\item[] {\bf Turyshev, S. G.,} {\it Motion of an extended
        bodies in the PPN formalism.}, Ph.D. dissertation, Moscow State
        University, Russia,   1990.
\item[] {\bf Tyler, G.L.,} ``Radio propagation experiments in the outer solar
        system with {\it Voyager}''. {\it Proc.\ IEEE} {\bf 75},
        1404-1431, 1987.
\item[] {\bf Tyler, G.L., Sweetnam, D.N.,  Anderson, J.D.,  Borutzki, S.E.,
        Campbell, J.K., \hyphenation{Es-hle-man}, V.R.,
        Gresh,  D.L.,  Gurrola, E.M.,
        Hinson, D.P., Kawashima,  N.,  Kursinski,  E.R.,  Levy, G.S.,
        Lindal,  G.F.,
        Lyons, J.R., Marouf, E.A.,  Rosen, P.A.,  Simpson, R.A., and
        Wood,  G.E.,}
        ``{\it Voyager}  radio science observations of Neptune and Triton''.
        {\it Science}, {\bf 246}, 1466-1473, 1989.
\item[] {\bf Tyler, G.L., Balmino,  G.,  Hinson, D.P.,  Sjogren, W.J.,
        Smith, D.E.,
        Woo,  R., Asmar, S.W.,  Connally, M.J., Hamilton, C.L., and
        Simpson, R.A.,} ``Radio science investigations with {\it Mars
Observer}''.
        {\it J. Geophys.\ Res.}, {\bf 97}, 7759-7779, 1992.
\item[] {\bf Vessot, R. F. C.  et al.,} ``Test of Relativistic Gravitation
        with a Space-Borne Hydrogen Maser''. {\it Phys. Rev. Lett.},
        {\bf 45}, 2081-2084, 1980.
\item[] {\bf Vincent, M.A., and  Bender, P.L.,} ``Orbit Determination and
        Gravitational Field Accuracy for a Mercury Transponder Satellite''.
        {\it J. Geophys.  Res.}, {\bf 95}, 21,357-21,361, 1990.
\item[] {\bf Williams, J. G., Newhall X X and  Dickey, J. O.,}
        Submitted to  {\it Phys. Rev. D},  1995.
\item[] {\bf Will, C.M.,}
        ``Theoretical framework for testing relativistic gravity. III.
        Conservation laws, Lorentz invariance, and values of the
        PPN parameters.''
        {\it ApJ}, {\bf 169}, 125-40,  1971.
\item[] {\bf Will, C. M. and Nordtvedt, K.,}
        ``Conservation laws and preferred frames in relativistic gravity. I.
        Preferred frame theories and an extended PPN fromalism''.
        {\it ApJ}, {\bf 177}, 757-74, 1972.
\item[] {\bf Will, C. M.,} {\it Theory and Experiment in Gravitational
Physics.}
        (Rev. Ed.)  Cambridge U. Press, Cambridge, England, 1993.
\item[] {\bf Ulrich, R. K.,} ``The influence of partial ionization
        and scattering states on the solar interior structure''.
       {\it ApJ}, {\bf258}(1), 404-413,  1982.
\end{itemize}

{\Large\bf Appendix A: The astrophysical parameters
used in the paper.}

$$\hbox{Solar radius}:  R_\odot  = 695,980 \hbox{ km},$$
$$\hbox{Solar gravitational constant}:  \mu_\odot  =
{G M_\odot\over c^2}=1.4766 \hbox{ km},$$
$$\hbox{Solar quadrupole coefficient (Brown {\it et al.}, 1989)}:
  J_{2\odot} = (1.7\pm0.17)\times 10^{-7},$$
$$\hbox{Solar rotation period}:
\tau_\odot =  25.36 \hbox{ days},$$
$$\hbox{Mercury's mean distance}:
 a_M  = 0.3870984  \hbox{ AU}= 57.91\times 10^6 \hbox{ km},$$
$$\hbox{Mercury's radius}: R_M  = 2,439  \hbox{ km},$$
$$\hbox{Mercury's gravitational constant}:
 \mu_M = {G M_M\over c^2}= 1.695\times 10^{-7}\mu_\odot,$$
$$\hbox{Mercury's sidereal period}:
 T_M  = 0.241  \hbox{ yr} =87.96 \hbox{ days},$$
$$\hbox{Mercury's rotational period}: \tau_M  =  59.7 \hbox{ days },$$
$$\hbox{Eccentricity of Mercury's orbit}:
e_M  =  0.20561421,$$
$$\hbox{Jupiter's gravitational constant}:
 \mu_J = 9.547\times 10^{-4}\mu_\odot,$$
$$\hbox{Jupiter's sidereal period}:
T_J  = 11.865  \hbox{ yr},$$
$$\hbox{Astronomical Unit}:
 AU = 1.49597892(1)\times 10^{13}  \hbox{ cm},$$

\section{Tables}
\begin{center}
\begin{tabular}{clc}
\multicolumn{3}{c}{\bf Table 1}\\
\multicolumn{3}{c}{Mercury Radar-Ranging Measurements}\\
\hline
\bf Timespan & \bf Antenna & \bf Number\\
{} & {} & \bf of Observations\\
\hline
1967-1971 & Arecibo & 85\\
1966-1971 & Haystack & 217\\
1971-1974 & Goldstone & 38\\
1974-1975 & Mariner 10 & 2\\
1978-1982 & Arecibo & 157\\
1986-1990 & Goldstone & 132\\
\hline
\end{tabular}
\end{center}

\end{document}